\documentclass[aps,prl,twocolumn, superscriptaddress,showpacs]{revtex4-1}

\usepackage{amsmath}
\usepackage{graphicx}
\usepackage{amssymb}
\usepackage{amsthm, amscd}
\usepackage{amsfonts}
\usepackage{subfigure}
\usepackage{pstricks}
\usepackage{epsfig}
\usepackage{epstopdf}
\usepackage{hyperref}
\usepackage{bm}

\newcommand{\dslash}{\not{\hbox{\kern-2pt $\partial$}}}

\newcommand{\mat}[4]{\left(\begin{smallmatrix}#1 & #2\\#3 & #4\end{smallmatrix}\right)}

\newcommand{\bb}{\vec{b}}
\newcommand{\dd}{\phi_d}

\newcommand{\KK}{\vec{K}}
\newcommand{\Or}{\bm{O}_R}
\newcommand{\Ol}{\bm{O}_L}
\newcommand{\Ir}{\bm{I}_R}
\newcommand{\Il}{\bm{I}_L}

\newcommand{\Ssc}{S_{\text{scatt}}}
\newcommand{\Stot}{S}

\begin{document}


\title{Straining the identity of Majorana fermions}
\author{Andrej Mesaros}
\affiliation{Instituut-Lorentz for Theoretical Physics, Universiteit Leiden, P. O. Box 9506, 2300 R A Leiden, The Netherlands}
\affiliation{LASSP, Physics Department, Clark Hall, Cornell University, Ithaca, NY 14853-2501}
\author{Stefanos Papanikolaou}
\affiliation{LASSP, Physics Department, Clark Hall, Cornell University, Ithaca, NY 14853-2501}
\author{Jan Zaanen}
\affiliation{Instituut-Lorentz for Theoretical Physics, Universiteit Leiden, P. O. Box 9506, 2300 R A Leiden, The Netherlands}
\date{\today}
\begin{abstract}
We propose an experimental setup of an interferometer for the observation of 
neutral Majorana fermions on topological insulator --- superconductor --- ferromagnet junctions. We show that the extended lattice defects naturally present in materials, dislocations, induce spin currents on the edges while keeping the bulk time-reversal symmetry intact. We propose a simple two terminal conductance measurement in an interferometer formed by two edge point contacts, which reveals the nature of Majorana states through the effect of dislocations. The zero temperature magneto-conductance changes from {\it even} oscillations with period $\phi_0/2$  ($\phi_0$ is the flux quantum $h c/e$) to {\it odd} oscillations with period $\phi_0$, when non-trivial dislocations are present and the Majorana states are sufficiently strongly coupled. Additionally, the conductance acquires a notable asymmetry as a function of the incident electron energy, due to the topological influence of the dislocations, while resonances appear at the coupling energy of Majorana states.
\end{abstract}

\maketitle

There is a strong interest to realize, observe and manipulate Majorana fermions, because of the non-Abelian statistics they possess~\cite{moore91}, being the basis for topological quantum computation~\cite{kitaev03}. Majorana fermions have been argued to be present in the $\nu=5/2$ fractional quantum Hall state~\cite{moore91,read00}, in the $p-$wave superconductor Sr$_2$RuO$_4$~\cite{dassarma06} and in topological insulator-superconductor junctions~\cite{fu08,*fu09,nilsson08}. Topological insulators (TI)~\cite{bernevig06, *murakami06, *fu07,*qi08} have gapless edge (2DTI) or surface (3DTI) states that are helical and topologically protected in the absence of time-reversal symmetry (TRS) breaking fields. Breaking TRS by depositing an insulating magnetic (M) material can open an energy gap leading to a novel surface quantum Hall effect with $ \sigma_{\rm xy}= \pm e^2/2h$~\cite{fu07b}. Further, deposition of a superconductor (S) on the edge or surface leads to Majorana
bound states (MBS) at an S-TI-M interface where the gap changes sign~\cite{fu08}.
These Majorana fermions have proven to be very elusive since they are neutral, and there are a few proposals for their observation, ranging from rather indirect tunneling experiments~\cite{bolech07,*semenoff07,*tewari08}, to  interference experiments~\cite{fu09b,Benjamin:2010p3417,Lutchyn}. In this paper, we make a step further and propose a novel interferometer which preserves TRS and can be used not only for identifying the Majoranas, but also for the dual purpose of understanding the fundamental properties of topological lattice defects.

\begin{figure}[t!]
\centering
\includegraphics[width=0.45\textwidth]{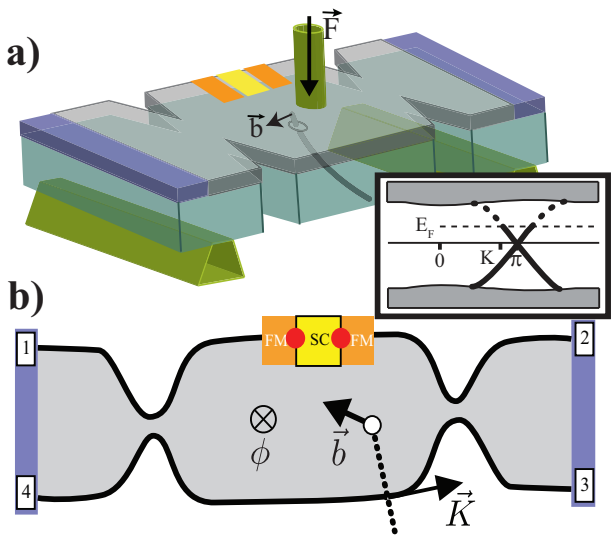}
\caption{Experimental setup for observing neutral Majorana fermion bound 
states (MBS). a) Three-point bending (by force $\vec{F}$), moving an 
example dislocation line (Burgers vector $\bb$) which pierces the 
substrate (green) and the 2D Topological Insulator (TI, grey) at the 
circle. At one TI edge there is a pair of superconductor (SC, yellow 
strip) - ferromagnet (FM, orange strip) - TI junctions. Contacts are 
marked violet. b) Edge modes of the 2DTI (grey area) traverse the 
interferometer, where the MBS (red dots) are present in the upper arm.
  The dislocation induces a translation on the half-plane of missing 
atoms (dotted line), causing a topological phase shift 
$\exp{(i\KK\cdot\bb)}$, where $\KK$ is the three-dimensional embedding 
of the edge Fermi momentum $K$. Inset: Schematic 2DTI band structure 
showing $K$ near $\pi$.
}
\label{fig:1}
\end{figure}

We propose a standard Aharonov-Bohm (AB) interferometer (Fig.~\ref{fig:1}b), where the presence of dislocations within the interferometer area causes a topological phase shift on the edge states due to the translational effect of the dislocation Burgers vector on the edge wavefunction. This AB effect~\cite{Mesaros:2009p1575} is analogous to the effect of pierced magnetic flux~\cite{fu09b}, except that it {\it preserves} TRS. The magnetic flux induces electrical current flow, the persistent current, in the ground state.
 Analogously, dislocations induce the TRS invariant counterpart, {\it dissipationless spin currents}. Spin currents are typically hard to observe, but appear useful for MBS detection. Dislocations in 3DTI were also found to host interesting states~\cite{Ran:2009p1115,*RanarXiv1006.5454R}.

\begin{figure}[t]
\centering
\includegraphics[width=0.45\textwidth]{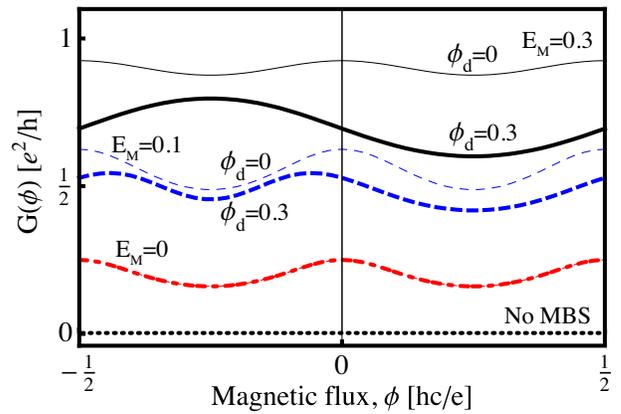}
\vspace{-2mm}
\caption{Magnetoconductance $G(\phi)$ at zero energy (voltage) as function of dislocations and Majorana couplings. In absence of MBS, $G(\phi)$ vanishes (dotted black line). The dash-dotted (red), dashed (blue) and full (black) curves correspond to three regimes of MBS coupling $E_M=0,0.1,0.3$, respectively, with respect to MBS-edge coupling $\Gamma=0.1$ (units $\hbar v_F/L$). The case of absence ($\dd=0$) or presence ($\dd=0.3$) of dislocations is distinguished by thin and thick lines, respectively, for each $E_M$ value. Without dislocations, the result $\delta G\sim\cos{(2\phi)}$ is robust to interferometer parameter changes. The presence of dislocations affects {\it only} the cases of coupled MBS, $E_M\neq 0$, by suppressing the $\phi_0/2$ harmonic in $G(\phi)$ in favor of the $\phi_0$, which is always {\it odd}, i.e. $\sin{(\phi)}$.}
\label{fig:4}
\end{figure}

The STIM interface
locally breaks TRS and particle-hole symmetry (PHS)~\cite{fu08, *fu09, nilsson08}, so that 
clear experimental signatures in the two-terminal AB interferometer are expected, e.g.
asymmetry of the magnetoconductance ($G(\phi)\neq G(-\phi)$, where $\phi$ is the threaded magnetic flux), being typically absent due to TRS.
 We find that magnetoconductance remains {\it even} in the presence of MBS, due to the topological {\it helicity} symmetry (exchange of the left/right moving up/down spin, for the left/right moving down/up spin edge modes). But when dislocations are present (this is controlled by straining the bulk of the TI, Fig~\ref{fig:1}a), a spin current is introduced in the interferometer which is sensitive to the helicity flip and therefore can detect the signatures of MBS. Most strikingly, the oscillations $\delta G(\phi)$ switch from {\it even} with period $\phi_0/2$  ($\phi_0$ is the flux quantum $h c/e$) to {\it odd} oscillations with period $\phi_0$ when dislocations enter the device and the MBS are coupled (Fig.~\ref{fig:4}), while oscillations vanish in absence of MBS at the STIM. We predict that the conductance satisfies 
 \begin{eqnarray}
 G(\phi,E,\phi_d)=G(-\phi,-E,-\phi_d)
 \label{eq:0}
 \end{eqnarray}
 ($E$ incident electron energy, and $\dd$ the dislocation scattering topological phase), which allows the use of the topological effect of dislocations ($\dd$) as a new control parameter (absent in all existing proposals~\cite{fu09b,Benjamin:2010p3417,Lutchyn}) to bring out the signatures of the MBS.

In the example of 2DTI realized in Hg(Cd)Te quantum wells~\cite{bernevig06,konig07}, dislocations seem neatly controllable. According to detailed structure studies as in Ref.~\cite{hgte1985}, at low temperatures and at the yield stress of $\simeq10-100\text{MPa}$, there are $10^{-10}\text{m}^{-2}$ dislocations, giving $\lesssim 1$ mobile dislocations piercing a $5\mu\text{m}\times 5\mu\text{m}$ sample. After yielding, no additional stress is needed in Hg(Cd)Te to move defects, so they move freely and independently. Upon reaching a high $10\%$ strain, with a total Burgers vector of $2*10^{-9}\text{m}$, one expects that $\simeq150$ dislocations have passed (glided) through such a sample. With dislocations being the most natural and abundant topological defects in crystals,
  we further expect the rightful use of dislocation induced spin currents as novel TRS probes in the future.

Our interferometer is made of a 2DTI shaped by two point contacts (Fig.~\ref{fig:1}b), and we model it using the Landauer-B\"uttiker scattering matrix formalism valid at low temperatures in the regime of coherent transport~\cite{Roth:2009p3241}. Edge segments comprising the interferometer support one electron and one hole {\it chiral} mode. The Bogoliubov - de Gennes Hamiltonian describing each edge segment is,
\begin{equation}
  \tau_3(v_F \hat{p}\;\sigma_3+A_d+\tau_3\sigma_3eA/\hbar c-E_F)\Psi=E\Psi,
  \label{eq:1}
\end{equation}
where $\hat{p}\equiv -i\hbar\partial/\partial x$, $E_F$ the Fermi energy, $v_F$ the Fermi velocity, and $A$ the magnetic vector potential, and the $x$-axis is along the given edge segment. The four-component spinor is $\Psi=(\Psi_{e\uparrow},\Psi_{e\downarrow},\Psi_{h\uparrow},\Psi_{h\downarrow})^T$ while the $\tau$ matrices mix the electron and hole parts of the wavefunction, and $\sigma$ the spin components.

The effect of dislocations is contained in the potential $A_d$ of Eq.~\eqref{eq:1}. It encodes for the AB effect $\exp{(\frac{i}{\hbar v_F}\oint A_d\;dx)}=\exp{(i 2\pi \dd)}$, with pseudo-flux $\dd$ stemming from the topological effect of the dislocation on the wavefunctions on the edge. It is well known that this effect is described by 
a {\it translation} by the Burgers vector $\bb$ on traversal of electron around the dislocation core line threading the TI inside the ring-shaped area of the interferometer~\cite{Kleinert}. The translation operator $\exp{(i\KK\cdot\bb)}\equiv\exp{(i 2\pi \dd)}$ is determined by the three-dimensional Burgers vector $\bb$ of the dislocation line that can be of any type (edge, screw or mixed). The vector $\KK$ is the three-dimensional embedding of the edge wavefunction wavevector $K$~\cite{fu07} (see Fig.~\ref{fig:1}a). The dislocation effects discussed in this paper depend on one-dimensional momentum $K$ on the edge being non-zero; such 2DTI variety is not yet observed, but it could exist in Hg(Cd)Te~\cite{Dai:2008p4596}, or Heusler alloys~\cite{heusler10}.
 Dislocations preserve time reversal and particle hole (PHS) symmetries, represented by $T=i\sigma_2C$ and $\Xi=\tau_2\sigma_2C$, respectively, with $C$ the complex conjugation, and they are distinct from ordinary disorder due to their intrinsic gauge symmetry. Generically, as in HgTe wells, edge segments do not exhibit PHS, so we checked that our results are robust to breaking PHS by assigning different velocities to edge states below and above $E=0$ (e.g. the edge energy spectrum observed in 3DTI~\cite{Xia:2009p4110}).

{\it Scattering formalism--} The Hamiltonian of Eq.~\eqref{eq:1} determines the energy dependent wavevector of the left (spin down on upper edge), and right (spin up on upper edge), moving electron, as well as their time reversed hole pairs.
The point contacts, the two halves of the upper ring arm, the coupled MBS between the two upper arm halves, and the lower ring arm are all described by single scattering points with corresponding matrices ($\Ssc$). Each matrix $\Ssc$ connects the amplitudes ($\Ol$ and $\Or$) of the modes \emph{outgoing} to the left/right ($L$/$R$) side, to the amplitudes ($\Il$ and $\Ir$) of the \emph{incoming} modes, with respect to that particular scatterer.
Using $O^T=(\Ol,\Or)^T=(o^L_{e\downarrow},o^L_{h\uparrow},o^R_{e\uparrow},o^R_{h\downarrow})^T$ and $I^T=(\Il,\Ir)^T=(i^L_{e\uparrow},i^L_{h\downarrow},i^R_{e\downarrow},i^R_{h\uparrow})^T$ 
we have $O=\Ssc I$.
The matrices have a block structure $\Ssc=\mat{r}{t}{t'}{r'}$, representing reflection ($r$, $r'$) and transmission ($t$, $t'$), where each block has electron/hole components, {\it e.g.} $t=\mat{t^{\text{ee}}}{t^{\text{he}}}{t^{\text{he}}}{t^{\text{hh}}}$. When TRS is obeyed, backscattering is forbidden in the single-particle formalism, $r=r'=0$~\cite{Xu:2006p3246}.
We compute  the total scattering matrix $\Stot$ for the four leads~(labeled from $1$ to $4$, Fig.~\ref{fig:1}b), determining the conductance of the device.
\begin{figure}[t]
\centering
\includegraphics[width=0.45\textwidth]{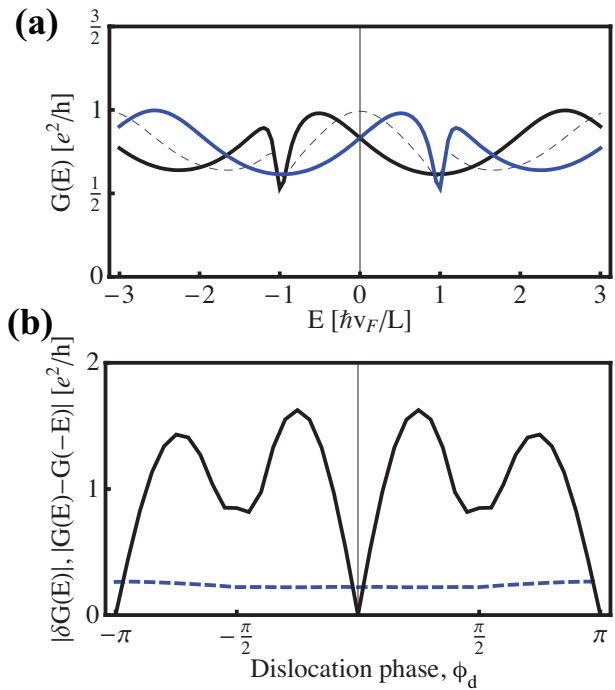}
\vspace{-2mm}
\caption{The energy (voltage) dependent conductance, as function of dislocation, with $E_M=1$. (a) Typical curves: no dislocation (dashed thin line); dislocation phase $\dd=0.1$ (thick black); and $\dd=-0.1$ (thick light grey (blue)). Note the resonances at $\pm E_M=\pm 1$, and the dislocation induced asymmetry. (b) The asymmetry of the $G(E)$ curves (full black line), calculated as ${\text Max_E}[G(E)-G(-E)]$ on the interval $E\in[0,3]$ (it reaches $2$ for purely odd $G\sim\sin{(E)}$). The dashed grey (blue) line shows the amplitude of $G(E)$ oscillations around the mean. The curves are robust to changes in $E_M$.
}
\label{fig:2}
\end{figure}

Particle conservation is enforced by $S^\dagger S=\openone$. For scattering matrices connecting two edge segments, TRS demands $\Ssc(\phi)=-\alpha_3\Ssc(-\phi)^T\alpha_3$, and PHS $\Ssc(E)=\beta_1\Ssc(-E)^*\beta_1$, where $\alpha$ and $\beta$ are Pauli matrices acting on the $L/R$ and $e/h$ indices of $\Ssc$, respectively. For scattering involving all four edges (like in $\Stot$) one should only replace $\alpha$ by $\alpha\otimes\alpha'$, where $\alpha'$ matrices exchange the two leads on the same side (i.e. $1$ and $4$, or $2$ and $3$). 
The scattering caused by the coupling to, and propagation through the two MBS in the upper arm is given by the scattering matrix $S_{MBS}$ found in Ref.~\cite{nilsson08}. It is determined by two energy scales, the coupling between the two MBS $E_M$, and the coupling of edge states to the MBS  $\Gamma$.
Length is measured in units of the ring circumference $L$, $\phi$ in units of the flux quantum $\phi_0=h c/e$, and energy in units of $\hbar v_F/L$.
We consider the scattering mechanisms as follows: 
(a) Propagation in the lower arm $S_{\text{low}}$ is determined by nonzero elements $t^{ee}_{\text{low}}=\exp{[i l_d(E+2\pi\dd-2\pi\phi)]}$, $t^{hh}_{\text{low}}=\exp{[i l_d(E-2\pi \dd+2\pi\phi)]}$, where $l_u$ is the length of the lower arm; 
(b) In the upper arm segments $S_{\text{up}} =S_{\text{low}}^T$, with $l_d$ replaced by $l_{u1}$, and $l_{u2}$ in the two segments, respectively; 
(c) Without loss of generality we take the point contact scattering matrix $S_{\text{PC}}$ to be real and satisfying the TRS and PHS symmetries~(edge segments are ordered as $(1,4)$ on the left and $(2,3)$ on right, cf. Fig.~\ref{fig:1}):
\begin{equation}
  \label{eq:2}
  S_{\text{PC}}=\begin{pmatrix}
0 & a & b & b' \\
a & 0 & -b' & b \\
b & b' & 0 & -a \\
-b' & b & -a & 0
\end{pmatrix}\otimes\beta_0,
\end{equation}
with $\beta_0=\openone$, and $a^2+b^2+b'^2=1$.
Parameter $a$ describes the coupling of the ring-shaped middle of the interferometer to the leads ($a=0$ corresponds to an isolated ring with $G=0$). The ratio $\epsilon\equiv b/b'$ measures the asymmetry of current injected into the lower and upper ring arms~($\epsilon=0$ corresponds to all particles from lead $1$ being injected into the lower arm, and all from lead $4$ into the upper). Following Refs.~\cite{Hou:2009p3262,*Teo:2009p3261}, in the present single particle scattering we attain the conductance of the charge-conductor/spin-insulator (CI) state by choosing $a=1/\sqrt{3}$, $\epsilon=1$, being in the regime of Luttinger liquid coupling $g_c>2$. In the realistic case of intermediate $0<a<1$, the dependence on $a$ and $\epsilon$ is weak, so we present results for CI contacts. The conductance is given by $G=e^2/h\sum_{i=1,4\atop {j=2,3}}\left(|S_{ij}^{ee}|^2-|S_{ij}^{he}|^2\right)$, where $i,j$ label the leads, and holes contribute opposite charge current from electrons. 
The zero temperature conductance at zero voltage corresponds to taking $E=0$, while at low temperature and voltage difference, $E$ is given by the external voltage ($E=e V_1$).
We consider $E_F=0$ (small $E_F$ is negligible when $K\simeq\pi$, see inset Fig.~\ref{fig:1}), and fix $l_{u1}=l_{u2}=l_d/2=L/2$, while the results are insensitive to the asymmetry in $l_{u1}$ and $l_{u2}$. The point contact parameters $a$, $\epsilon$ are set to be the same in the left and right contact, since results are robust to this asymmetry too. 

The symmetry expressed in Eq.~\eqref{eq:0}  is most revealing since it controls the behavior of the conductance  $G(\phi,E)$, given the changes in the net Burgers vector $d$. It represents the invariance of the edge states to switching the spin orientation of left and right moving carriers.~(This orientation is set by the sign of the bulk spin-orbit coupling.) For the scattering on the edge, this switch is represented by conjugation $C$, i.e. the combined time-reversal and spin-flip operation. In this case, it follows that $\Ssc(E)=\Ssc(-E)^*$ and non-trivially for the case of $S_{MBS}$, this property holds because $H_M^*=-H_M$. The two-level Hamiltonian $H_M$ fundamentally obeys the relation because the Majorana fields are real~($\gamma_a^\dagger=\gamma_a$). We expect the spin-flip symmetry to be robust in absence of Zeeman type coupling to out-of-plane magnetic fields.

We first consider the effect of dislocations on a trivial interferometer, one without a STIM interface. The presence of $\dd\neq 0$ introduces a deviation from evenness in $G(E)$, as the symmetry $G(E,\dd)=G(-E,-\dd)$ suggests. The magnetoconductance $G(\phi)=G(-\phi)$ stays even, protected by TRS in a two terminal measurement. However, the nature of the $G(\phi)$ oscillations switches from dominantly universal conductance fluctuations (UCF), i.e. period $\phi_0$, to a dominantly period $\phi_0/2$ nature, when dislocation is introduced. 

Secondly, we introduce the STIM interface into the upper arm of the interferometer (cf. Fig.~\ref{fig:1}). If there are {\it no} MBS forming, the STIM is a segment of gapped edge states with a TRS violation.
   The absence of MBS is modeled by setting $\Gamma=0$ (decoupling from the edges). In this case, the magnetoconductance oscillations $\delta G(\phi)$ vanish. The dislocations influence the oscillations, and $G(E,\dd)=f(E-\dd)$, with $f(x)=b'^2(1-a^4)\left[(1+a^2)(1+\epsilon^2)+4a\epsilon\cos{(x/2)}\right]/\left[1+a^8+2a^4\cos{(2x)}\right]$ shows clearly that the asymmetry of $G(E)$ is controlled by the dislocations. The effect persists in the limit 
   where the central ring is decoupled from the leads ($a=0$): the spectrum of the ring is given by the solutions of $\cos{(2E)}=\cos{(\pi \dd)}$, and shows the symmetries $\{E_n(\dd)\}\neq-\{E_n(\dd)\}$ and $\{E_n(\dd)\}=-\{E_n(-\dd)\}$. 


The general asymmetry features in $G(E)$ due to dislocations persist when MBS are added, and new signature effects appear in the magnetoconductance as dislocations are manipulated.
The $G(E)$ shows oscillatory behavior, with resonances at $\pm E_M$, shown in Fig.~\ref{fig:2}(a). In Fig.~\ref{fig:2}(b), we provide a summary of the dislocation effect on the behavior of $G(E)$. Introduction of non-zero dislocation phase causes a large asymmetry that persists for all values of $E_M$. If the flux $\phi$ is present, $G(E)$ becomes asymmetric at any value of $\dd$, and more strongly as $E_M$ increases (note that when MBS are absent, there is no dependence on $\phi$). The last observation was made also for a more complicated hypothetical interferometer~\cite{Benjamin:2010p3417}.

Fig.~\ref{fig:4} presents the characteristic influence of dislocations and Majorana states on the magnetoconductance at zero energy ({\it i.e.}~zero voltage at low temperatures). As announced, even though TRS is broken by the MBS scattering, a resulting non-even $G(\phi)$ is observed only in the presence of dislocations. Namely, $\delta G(\phi)$ has two prominent Fourier components, and both have {\it definite parity}: the UCF in the form of $\sin{(\phi)}$, and the harmonic $\cos{(2\phi)}$. When $E_M=0$ (MBS decoupled from each other), the UCF vanish. However, when $E_M\neq 0$, dislocations show a clear signature: in their presence, as $E_M/\Gamma$ increases, the harmonic is suppressed in favor of the UCF, and therefore simultaneously the transformation from even to odd $G(\phi)$ is observed. If $E_M>\Gamma$, a small value of $\dd$ (e.g. $0.05$) already causes a linear $G(\phi)$ up to $|\phi|\lesssim 1/4$ (cf. Fig.~\ref{fig:4}).

In conclusion, we demonstrated the usefulness of shear stress manipulated dislocations in observing neutral Majorana fermions in TIs. We found clear signatures of dislocation - MBS interplay in magnetoconductance oscillations at zero energy, and showed the enhanced conductance symmetry of Eq.~\eqref{eq:0}, a direct consequence of the symmetry of the TI and the presence of dislocations.



\begin{acknowledgments}
We thank A. Vishwanath, J. Bardarson, A. Akhmerov and V. Juri\v{c}i\'{c} for useful discussions. AM is grateful for hospitality of E.-A. Kim. AM acknowledges financial support by the Nederlandse Organisatie voor Wetenschappelijk Onderzoek~(NWO), and SP by the DOE-BES through DE-FG02-07ER46393.
\end{acknowledgments}
\bibliography{majorana1.bib}
\end{document}